\documentclass[12pt]{article}
\usepackage{graphicx}
\usepackage{amsmath,amssymb}
\usepackage{epsfig}
\usepackage{latexsym}
\newcommand {\ga} {\ {\raise-.5ex\hbox{$\buildrel>\over\sim$}}\ }
\newcommand {\la} {\ {\raise-.5ex\hbox{$\buildrel<\over\sim$}}\ }

\newcommand{\ba}{\begin{eqnarray}} \newcommand{\ea}{\end{eqnarray}}
\newcommand{\be}{\begin{equation}} \newcommand{\ee}{\end{equation}}

\usepackage{amsmath,amssymb,amsthm,graphicx,bm,color}

\newcommand{\nn}{\nonumber}

\renewcommand{\Delta}{\varDelta}
\renewcommand{\Gamma}{\varGamma}
\renewcommand{\Omega}{\varOmega}
\renewcommand{\Phi}{\varPhi}
\renewcommand{\Psi}{\varPsi}
\renewcommand{\Sigma}{\varSigma}
\renewcommand{\Theta}{\varTheta}
\renewcommand{\epsilon}{\varepsilon}

\newcommand{\irrep}[1]{\ensuremath{\mathbf{#1}}}

\newlength{\irrepwidth}
\newlength{\irrepbarthickness}
\newlength{\irrepbarheight}
\newcommand{\irrepbar}[1]
{
    \ensuremath{
        \settowidth{\irrepwidth}{\ensuremath{\mathbf{#1}}}%
        \makebox[0pt][l]{\ensuremath{\mathbf{#1}}}%
        \rule[\irrepbarheight]{\irrepwidth}{\irrepbarthickness}
    }
}

\begin{document}

\vspace*{-0.5cm}
\begin{flushright}
OSU-HEP-09-05\\
\end{flushright}
\vspace{0.5cm}






\begin{center}
{\Large {\bf Natural fermion mass hierarchy and mixings in family unification} }\\

\vspace*{1.5cm} James B. Dent\footnote{Email address: jbdent@asu.edu}$^{,\dag}$,\footnote{Present address: Department of Physics and School of Earth and Space Exploration, Arizona State University, Tempe, AZ 85287-1404}$^{\*}$
Robert Feger\footnote{Email address: thomas.w.kephart@vanderbilt.edu}$^{,\ddag}$,
Thomas W. Kephart\footnote{Email address: robert.feger@vanderbilt.edu}$^{,\ddag}$,
and  S. Nandi\footnote{Email address: s.nandi@okstate.edu}$^{,\#}$\\

\vspace*{0.5cm}
$^{\dag}${\it Department of Physics and Astronomy, Vanderbilt University, Nashville, TN 37235}\\

$^{\#}${\it Department of Physics and Oklahoma Center for High Energy Physics,\\
Oklahoma State University, Stillwater, OK 74078}\\
\vspace*{0.5cm}\today
\end{center}


\begin{abstract}
We present an $SU(9)$ model of family unification with three light
chiral families, and a natural hierarchy of charged fermion masses
and mixings. The existence of singlet right handed neutrinos with
masses about two orders of magnitude smaller than the GUT
scale, as needed to understand the light neutrinos masses via the
see-saw mechanism, is compelling in our model.
\end{abstract}

\medskip



\section{Introduction}
The ideas of family symmetry and family unification has been with us
for a while. Grand unified theories lend themselves to construction
of such models,
but most of the early models
did not go as far as considering fermion masses and mixings. We
know that there is a five orders of magnitude hierarchy among the charged fermions
masses. There is also a two orders of magnitude hiearchy amongst the quark mixing angles.
In addition, there are strong suppressions for
the flavor changing neutral current processes.  With the fairly
accurate  data on charged fermion masses and quark mixings,
we are now in a position to attempt the construction of
family symmetry models that include these parameters. The
new data from the Tevatron and LHC  will  provide further
constraints on such family unified models.

We have studied a class of $SU(N)$ family unification models, i.e.,
models where the families are not due to simple replication of the
representation of the first fermion family. Grand unification
requires at least an $SU(5)$ gauge group, but the only reasonable
choice (that avoids exotic fermions) of representations for the
families are $(\irrep{10}+\irrepbar{5})_F$, and family unification is
impossible. In $SU(N)$ models, if the fermions all reside in totally
antisymmetric irreducible representations (irrep), then
there are guaranteed to be no exotic fermions. Such an idea was first
proposed by H. Georgi \cite{georgi}, and subsequently used by many authors
to build models with three chiral families \cite{Frampton:1979fd,
Frampton:1982pf,Frampton:1983pf,Frampton:2009ce}.
We write the $k^{th}$
totally antisymmetric irrep as $[{\bf 1}]^k$.  In $SU(N)$ there
are two invariant tensors from which we can construct group singlets
from the $[{\bf1}]^k$s. They are the Kroneker
$\delta^{\alpha}_{\beta}$ and the Levi-Civita tensor
$\epsilon_{\alpha_{1}\alpha_{2}...\alpha_{N}}$ or it's dual with all upper
indices. The indices $\alpha, \beta $, etc., run from 1 through $N$ of
$SU(N)$.

The number of totally antisymmetric irreps in the groups $SU(6)$ and $SU(7)$ are too small to
arrange the realistic mass and mixing  relations of the type to follow. An $SU(8)$ model of family unification has been proposed by S. Barr
\cite{Barr:2008gz,Barr:2008pn},
 but we find it possible to arrange a more detailed phenomenology in $SU(9)$, and this justifies our choice of gauge group.

\section{Three families from $SU(9)$}

In $SU(9)$ our three family representation is \cite{3fam}
$$
\irrepbar{126} + \irrep{84} +2(\irrep{36}) + 14(\irrepbar{9}).
$$
Note that this assignment is anomaly free.

We can write this in a shorthand notation as
$$
F^5 + F^3 + 2F^2 + 14F^8
$$
where we have made the replacement $[{\bf1}]^k\rightarrow F^k$.
 This can be written more
completely as
$$
{{\cal F}{{=}}{F}_{{\alpha}{\beta}{\gamma}{\delta}}{{+}}
{F}^{{\alpha}{\beta}{\gamma}}{{+}}{2}{{(}}{F}^{{\alpha}{\beta}}{{)}}
{{+}}{{14}}{{(}}{F}_{{\alpha}}{{)}}}
$$
where  symbolically, $F_{9-n} = \epsilon_{_{9}} F^n$.

  Now let us consider the breaking of the $SU(9)$ gauge symmetry to $SU(5)$, which can
  be done most simply with vacuum expectation values (VEVs) for a set of Higgs fundamentals,
  $H^{\alpha}_i$,
  $i=1,2,...$. Four successive VEV is sufficient to break $SU(9) \rightarrow SU(5)$. In this
  case the fermion irreps decompose as

$$\irrepbar{126} \rightarrow \irrep{5} + 4(\irrep{10}) + 6(\irrepbar{10}) + 4( \irrepbar{5}) + \irrep{1}$$
$$\irrep{84}     \rightarrow (\irrepbar{10}) + 4(\irrep{10}) + 6(\irrep{5}) +4(\irrep{1})$$
$$\irrep{36}     \rightarrow \irrep{10} + 4(\irrep{5}) +6(\irrep{1})$$
and
$$\irrepbar{9}   \rightarrow \irrepbar{5} +4(\irrep{1}).$$
Hence the complete set of fermions in the model is
$${\cal F}{{=}}3(\irrep{10}+\irrepbar{5})_F + 15(\irrep{5}+\irrepbar{5})_F + 7(\irrep{10}+\irrepbar{10})_F + 73(\irrep{1})_F.$$


Note that the fermions in $15(\irrep{5}+\irrepbar{5})_F + 7(\irrep{10}+\irrepbar{10})_F + 73(\irrep{1})_F$ all acquire masses at the
unification scale, leaving three massless chiral families in $3(\irrep{10}+\irrepbar{5})_F$.

\section{Fermion Masses and Mixings}

The assignment of the three light chiral families in the $SU(9)$
multiplets in our model are as follows.\\

3rd family: $ \irrepbar{126}_{3F} \rightarrow t_L,  t_R, b_L;   \irrepbar{9}_{3F} \rightarrow b_R$\\

2nd family: $\irrep{84}_{2F}      \rightarrow c_L,  c_R, s_L;   \irrepbar{9}_{2F} \rightarrow s_R$\\

1st family: $\irrep{36}_{1F}      \rightarrow u_L, u_R, d_L;    \irrepbar{9}_{1F} \rightarrow d_R$.\\

\noindent
In addition, we use several Higgs representations
shown in Table 2 below.
  The remaining $\irrep{36}_F$ and eleven
$\irrepbar{9}_{F}$ for the fermions do not contain any of the
$SU(5)$ level chiral fermions.

The charged fermions will receive masses from the Yukawa
interactions with the Higgs multiplets which have electroweak VEVs.
Since the top quark has a mass at the EW scale, its Yukawa coupling
is of order one. So it is very reasonable to assume that only the
top quark has a dimension four Yukawa interaction, while the allowed
interactions of the lighter quarks and charged leptons are of higher
dimensions, suppressed by a parameter $\epsilon$. We will identify
this parameter $\epsilon$ with the ratio of the $SU(5)$ singlet
Higgs VEV, $<1>$ and the unification scale, $M$.

\subsection{Yukawa interaction for the up sector}
To achieve an acceptable mass spectrum, we need  a set of Higgs
fields and discrete symmetries that leads to a hierarchy of  mass
terms. To achieve this end we  introduce   the Higgs representations
$\irrep{9}_H$, $\irrep{36}_{iH}$, and $\irrep{315}_H$, where
$i=1,2,3,4$. We next impose discrete symmetries  $Z_2\times
Z_2'\times Z_2''\times Z_3$ on the entire Lagrangian. The charge
assignments for the $SU(9)$ fermions that lead to chiral $SU(5)$
fields are given in Table \ref{tab:fermionDiscrete}, and the charge
assignments for the Higgses are given in Table
\ref{tab:higgsDiscrete}.

\begin{table}[htp]
\begin{center}
\begin{minipage}[t]{.4\textwidth}
\caption{Discrete charge assignments for the $SU(9)$ fermions}
\begin{center}
\begin{tabular}{|c|cccc|}
\hline
irrep                 & $Z_2$ & $ Z_2'$ & $Z_2'' $ & $Z_3$      \rule{0pt}{2.3ex}\\ \hline
$\irrep{36}_{1F}$     &   1   &   -1    &  -1      & $\alpha$   \rule{0pt}{2.3ex}\\ \hline
$\irrep{84}_{2F}$     &  -1   &   -1    &  -1      & $\alpha$   \rule{0pt}{2.3ex}\\ \hline
$\irrepbar{126}_{3F}$ &  -1   &    1    &   1      & $\alpha$   \rule{0pt}{2.3ex}\\ \hline
$\irrepbar{9}_{1F}$   &   1   &   -1    &   1      & $\alpha^2$ \rule{0pt}{2.3ex}\\ \hline
$\irrepbar{9}_{2F}$   &  -1   &   -1    &   1      & $\alpha^2$ \rule{0pt}{2.3ex}\\ \hline
$\irrepbar{9}_{3F}$   &  -1   &    1    &   1      & 1          \rule{0pt}{2.3ex}\\ \hline
\end{tabular}
\end{center}
\label{tab:fermionDiscrete}
\end{minipage}
\hspace{5mm}
\begin{minipage}[t]{.4\textwidth}
\caption{Discrete charge assignments for the Higgs fields}
\begin{center}
\begin{tabular}{|c|cccc|}
\hline
irrep             & $Z_2$ & $ Z_2'$ & $Z_2''$ & $Z_3$      \rule{0pt}{2.3ex}\\ \hline
$\irrep{36}_{1H}$ &  -1   &    1    &   1     & 1          \rule{0pt}{2.3ex}\\ \hline
$\irrep{36}_{2H}$ &  -1   &   -1    &  -1     & $\alpha^2$ \rule{0pt}{2.3ex}\\ \hline
$\irrep{36}_{3H}$ &   1   &   -1    &   1     & $\alpha^2$ \rule{0pt}{2.3ex}\\ \hline
$\irrep{36}_{4H}$ &   1   &    1    &   1     & $\alpha^2$ \rule{0pt}{2.3ex}\\ \hline
$\irrep{9}_H$     &  -1   &    1    &   1     & $\alpha$   \rule{0pt}{2.3ex}\\ \hline
$\irrep{315}_H$   &   1   &    1    &   1     & $\alpha$   \rule{0pt}{2.3ex}\\ \hline
\end{tabular}
\end{center}
\label{tab:higgsDiscrete}
\end{minipage}
\end{center}
\end{table}%

\pagebreak

Consistent with the $SU(9)\times Z_2\times Z_2'\times Z_2''\times Z_3$ symmetry, the
allowed Yukawa interactions for the up sectors are as follows.\\

Dimension 4:
$$ h^{u}_{33}\irrepbar{126}_{3F}\irrepbar{126}_{3F}\irrep{\bf{315}}_H,$$
where the $\irrep{315}$ is defined through $\irrep{9} \times  \irrepbar{36} = \irrepbar{9}+\irrep{315}$.\cite{315}\\

Dimension 5: \\

$\frac{1}{M} h^u_{32}$ $\irrepbar{126}_{3F}\irrep{84}_{2F}\irrep{36}_{2H}\irrepbar{9}_H ;$\\

 $\frac{1}{M} h^u_{22}$ $\irrep{84}_{2F}\irrep{84}_{2F}
( a_1 \irrep{36}_{4H}\irrepbar{315}_H + a_2 \irrep{36}_{1H}\irrep{9}_H ).$\\

Dimension 6:\\

$\frac{1}{M^2} h^u_{31}$ $\irrepbar{126}_{3F}\irrep{36}_{1F} (\irrep{36}_{1H})^2 \irrepbar{36}_{2H};$\\

$\frac{1}{M^2} h^u_{21}$ $\irrep{84}_{2F}\irrep{36}_{1F}
\{\irrepbar{9}_{1H}[a_3(\irrepbar{36}_{2H})^2
+a_4(\irrepbar{36}_{3H})^2+a_5(\irrepbar{36}_{4H})^2]+a_6\irrepbar{36}_{1H}(\irrepbar{315}_H)^2\}.$\\

Dimension 7:\\

$\frac{1}{M^3} h^u_{11}$ $\irrep{36}_{1F}\irrep{36}_{1F}\irrep{315}_H  [a_7(\irrep{315}_H)^3 + a_8(\irrep{36}_{4H})^3 + a_9 \irrep{36}_{4H}(\irrep{36}_{3H})^2 + a_{10}\irrep{36}_{4H}(\irrep{36}_{2H})^2],$\\

\noindent where the coefficients $a_K$ are all $O(1)$.
Note that the Yukawa interactions involving $\bar{c}_L  t_R$ have the same structure with $h^u_{32}$ replaced by $h^u_{23}$ above, and
similarly for the $\bar{u}_L c_R$ terms. Also no lower dimensional Yukawa
interactions are allowed for each terms.

In each of the Higgs multiplets, there are $\irrep{5}_H$, $\irrepbar{5}_H$, and
$\irrep{1}_H$ under $SU(5)$. From each of the Yukawa interactions, we use
one electroweak VEV arising from either $\irrep{5}_H$, or $\irrepbar{5}_H$, and the rest from
singlets. Thus a Yukawa interaction of dimension $4+ n$ above will give
rise to the mass matrix elements of the form

$$(h^u_{ij} v) \bar{u}_{iL}u_{jR} (\epsilon^n),\qquad\text{with}\qquad\epsilon = \frac {<{\bf1}>}{M},$$

\noindent where $<{\bf1}>$ is the VEV of the $SU(5)$ singlet field contained in the
above $SU(9)$ Higgs representations, and $M$ is the $SU(9)$ unification scale.

Collecting terms from the above Yukawa interactions, we obtain the following up quark
mass matrix:
\begin{eqnarray}
M_u = \left(
\begin{matrix}
h_{11}^u \epsilon^3 & h_{12}^u \epsilon^2 & h_{13}^u \epsilon^2 \cr
h_{21}^u \epsilon^2 & h_{22}^u \epsilon^1 & h_{23}^u \epsilon^1 \cr
h_{31}^u \epsilon^2 & h_{32}^u \epsilon^1 & h_{33}^u
\end{matrix}
\right)v.
\end{eqnarray} \nn

\subsection{Yukawa interaction for the down sector}

Since the bottom quark mass is very small compared to the EW scale,
we do not allow any dimension 4 Yukawa coupling in the bottom
sector. Again, consistent with the $SU(9)$ gauge symmetry and the
discrete symmetries, the allowed Yukawa interactions in the down
quark
sector are:\\[-1ex]

Dimension 4:\\[-1ex]

{\bf none}\\[-1ex]

Dimension 5:\\

$\frac{1}{M} h^d_{33}$ $\irrepbar{126}_{3F}\irrepbar{9}_{3F}
[b_1(\irrepbar{36}_{2H})^2
+b_2(\irrepbar{36}_{3H})^2+b_3(\irrepbar{36}_{4H})^2];$\\

Dimension 6:\\

$\frac{1}{M^2} h^d_{32}$ $ \irrepbar{126}_{3F}\irrepbar{9}_{2F} [b_4 \irrep{36}_{1H}
\irrep{36}_{3H}\irrep{9}_{H} +b_5\irrep{36}_{3H}\irrep{36}_{4H}\irrepbar{315}_H
+b_6\irrepbar{36}_{3H} (\irrep{315}_H)^2]$\\

$\frac{1}{M^2} h^d_{23}$ $\irrep{84}_{2F}\irrepbar{9}_{3F}[b_7\irrepbar{36}_{2H} \irrepbar{9}_H\irrepbar{315}_H +b_8 \irrep{36}_{1H}\irrepbar{36}_{2H}\irrepbar{36}_{4H}];$\\

$\frac{1}{M^2} h^d_{22}$ $ \irrep{84}_{2F}\irrepbar{9}_{2F}[b_9 \irrep{36}_{2H}\irrepbar{36}_{1H}\irrepbar{36}_{3H}+b_{10}\irrep{36}_{3H}\irrepbar{36}_{1H}\irrepbar{36}_{2H}].$\\

Dimension 7:\\

$\frac{1}{M^3} h^d_{31}$ $\irrepbar{126}_{3F}\irrepbar{9}_{1F}\{\irrep{36}_{1H}\irrepbar{36}_{3H} [b_{11}(\irrepbar{36}_{2H})^2+b_{12}(\irrepbar{36}_{3H})^2+b_{13}(\irrepbar{36}_{4H})^2]
+b_{14}\irrepbar{36}_{3H}\irrepbar{36}_{4H}\irrepbar{315}_{H}\irrepbar{9}_{H}\};$\\

$\frac{1}{M^3} h^d_{21}$ $\irrep{84}_{2F}\irrepbar{9}_{1F}\irrepbar{36}_{2H}\irrepbar{36}_{3H} (\irrepbar{315}_H)^2;$\\

$\frac{1}{M^3} h^d_{13}$ $\irrep{36}_{1F}\irrepbar{9}_{3F}\{\irrep{36}_{2H}\irrep{36}_{4H}[b_{15}(\irrep{36}_{2H})^2+b_{16}(\irrep{36}_{3H})^2+b_{17}(\irrep{36}_{4H})^2]
+b_{18}\irrep{36}_{2H}(\irrep{315}_H)^3\};$\\

$\frac{1}{M^3} h^d_{12}$ $\irrep{36}_{1F}\irrepbar{9}_{2F}[\irrepbar{315}_H (b_{19}\irrep{36}_{2H}\irrepbar{36}_{3H}\irrepbar{36}_{4H}+b_{20}\irrep{36}_{3H}    \irrepbar{36}_{4H}\irrepbar{36}_{2H}\\
~~~~~~~~~~~~~~~~~~~~~~~~~~~~~~~~~~~~~+b_{21}\irrep{36}_{4H}\irrepbar{36}_{2H}\irrepbar{36}_{3H})+b_{22}\irrep{9}_H\irrep{36}_{1H}\irrepbar{36}_{2H}\irrepbar{36}_{3H}];$\\

$\frac{1}{M^3} h^d_{11}$ $\irrep{36}_{1F}\irrepbar{9}_{1F}\irrep{36}_{1H}\irrep{36}_{2H}\irrep{36}_{3H}\irrep{36}_{4H}$,\\

\noindent
where all the coefficients $b_J$ are $O(1)$.

From the above Yukawa interactions, we obtain the following down
quark mass matrix.
\begin{eqnarray}
M_d = \left(
\begin{matrix}
h_{11}^d \epsilon^3 & h_{12}^d \epsilon^3 & h_{13}^d \epsilon^3 \cr
h_{21}^d\epsilon^3 & h_{22}^d \epsilon^2 & h_{23}^d \epsilon^2 \cr
h_{31}^d \epsilon^3 & h_{32}^d \epsilon^2 & h_{33}^d \epsilon^1
\end{matrix}
\right)v~.
\end{eqnarray}

We choose all our Yukawa couplings, $h^u_{ij}$ and $h^d_{ij}$  of
$O(1)$. The hierarchy in the fermion masses and mixings arises from
the different degree of suppression coming from the ratio of the
VEVs, i.e. $\epsilon$. Hence, given the choice of particle content and symmetry, the
hierarchy is technically natural.

\section{Phenomenology}

\textbf{FCNC and Higgs Decays:} Note that the up quark and down
quark mass matrices in our model are identical to those obtained in
Lykken, Murdock and Nandi (LMN)\cite{Lykken:2008bw}\cite{Babu:1999me}.
So, as shown there, if we choose the
parameter $\epsilon$ to be $\sim1/50$, our model is in good agreement
with all the quark masses and CKM mixings. However, the crucial
difference is that in our present model the existence of three light chiral
families, as well their mass and mixing hierarchies has its origin
in a gauge family symmetry, $SU(9)$. Another important difference is
that our singlet Higgs fields have masses close to the GUT scale,
not  the EW scale. Thus the phenomenology of this model is very
distinct from the LMN model. A further difference with the LMN model
is in the flavor changing neutral current (FCNC) interactions. In our
case, because the singlet Higgs fields are very heavy (close to the
GUT scale), their VEVS do not contribute to the mass matrix
elements, or to the Yukawa coupling matrix elements. Thus, in our
model, the mass matrices and the Yukawa coupling matrices for the
up and down quark sector are proportional, and hence there
are no flavor changing neutral current interactions at the tree
level. Thus the predictions of our model for the flavor changing
neural current processes are the same as in the SM. The same is true
for Higgs boson decays.\\

\textbf{Neutrino Masses and Mixings:}   Because we need $\irrepbar{9}_F$
of $SU(9)$ to obtain $\irrepbar{5}_F$ of $SU(5)$ for the chiral fermion families,
$SU(5)$ singlet fermions are unavoidable.
Thus, the existence of singlet right handed (RH)  neutrinos are
required in our model, similar to a $SO(10)$ GUT, and contrary to
a $SU(5)$ GUT. These RH neutrinos get Majorana masses from the
$SU(5)$ singlet Higgs whose VEVs are about 50 times smaller than the
GUT scale, as needed to explain the hierarchy of quark masses
and mixings. Thus our model naturally explains why the mass scale of
the RH neutrinos are smaller than the $SU(9)$ GUT scale as needed to obtain
the light neutrino masses at the observed level via the see-saw
mechanism. Furthermore, the Dirac mass terms between the light
neutrinos and the heavy RH neutrinos occur via  dimension 4
operators at the tree level. Hence, in agreement with observation, there will not be  large
hierarchies among the light neutrino masses  or among the
neutrino mixing angles.\\



 \textbf{Acknowledgments:}
 We  thank Carl Albright for some very helpful remarks about the first version of this manuscript.
 We thank Z. Murdock for useful discussions.
The work of RF was supported by a fellowship
 within the postdoc-Programme of the German Academic Exchange Service (DAAD). TWK thanks the
 Aspen Center for Physics
 for  hospitality while this work was in progress. SN would like to thank the
 Fermilab Theoretical Physics Department for warm hospitality and support
 during the completion of this work.
The work of JBD and TWK was supported by US DOE grant
DE-FG05-85ER40226. The work of SN was supported in part by the US
Department of Energy, Grant Numbers DE-FG02-04ER41306 and
DE-FG02-ER46140.


\begin{thebibliography}{99}
\bibitem{3fam}
We note that there are three family $SU(9)$ models with fewer total fermions, such as the fermions in $\irrep{84} +9(\irrepbar{9})$, but they would not serve our purposes here.


\bibitem{georgi}
H. Georgi, Nucl. Phys. {\bf B156} (1979) 126.


\bibitem{Frampton:1979fd}
  P.~Frampton and S.~Nandi,
  Phys.\ Rev.\ Lett.\  {\bf 43}  (1979) 1460.








\bibitem{Frampton:1982pf}
  P.~H.~Frampton and T.~W.~Kephart,
  Phys.\ Rev.\ Lett.\  {\bf 48} (1982) 1237 .


\bibitem{Frampton:1983pf}
  P.~H.~Frampton and T.~W.~Kephart,
  Nucl. Phys. {\bf B211} (1983) 239.



\bibitem{Frampton:2009ce}
  P.~H.~Frampton and T.~W.~Kephart,
  Phys.\ Lett.\  B {\bf 681} (2009) 343
  [arXiv:0904.3084 [hep-ph]].


\bibitem{Barr:2008gz}
  S.~M.~Barr,
  Phys.\ Rev.\  D {\bf 78} (2008) 055008
  [arXiv:0805.4808 [hep-ph]].

\bibitem{Barr:2008pn}
  S.~M.~Barr,
  Phys.\ Rev.\  D {\bf 78} (2008)  075001
  [arXiv:0804.1356 [hep-ph]].

  \bibitem{315}
  Note that we need $315_H$ because a term of the form
$\irrepbar{126}_{3F} \irrepbar{126}_{3F} \irrepbar{9}_H$, does not
contribute to the  top mass, since at the $SU(5)$ level it does not contain
$\irrep{10}_{3F}\irrep{10}_{3F}\irrep{5}_H$. Hence,
our discrete symmetries  do not allow a $\irrepbar{126}_{3F} \irrepbar{126}_{3F} \irrepbar{9}_H$ term.
We thank Carl Albright for clarifying this point.

\bibitem{Lykken:2008bw}
  J.~D.~Lykken, Z.~Murdock and S.~Nandi,
  Phys.\ Rev.\  D {\bf 79} (2009) 075014
  [arXiv:0812.1826 [hep-ph]].


\bibitem{Babu:1999me}
  K.~S.~Babu and S.~Nandi,
  Phys.\ Rev.\  D {\bf 62} (2000) 033002
  [arXiv:hep-ph/9907213];
 Gian F. Giudice and Oleg Lebedev, Phys. Lett. B665:79-85, 2008
  [e-Print arXiv:0804.1753(hep-ph)].

\end{thebibliography}
\end{document}